\documentclass[10pt,journal,compsoc]{IEEEtran}

\usepackage{amsmath,amsfonts}
\usepackage{algorithmic}
\usepackage{algorithm}
\usepackage{array}
\usepackage[caption=false,font=normalsize,labelfont=sf,textfont=sf]{subfig}
\usepackage{textcomp}
\usepackage{stfloats}
\usepackage{url}
\usepackage{verbatim}
\usepackage{graphicx}
\usepackage{cite}
\usepackage{booktabs} 
\usepackage{tabularx} 
\usepackage{threeparttable}
\usepackage{color}
\usepackage[normalem]{ulem}

\begin{document}
	%
	\title{GaussEdit: Adaptive 3D Scene Editing with Text and Image Prompts}
	%
	%
	%
	%
	\author{Zhenyu~Shu,
		Junlong~Yu*,
		Kai~Chao,
		Shiqing~Xin,
		Ligang~Liu
		\IEEEcompsocitemizethanks{
			\IEEEcompsocthanksitem Zhenyu Shu is with School of Computer and Data Engineering, NingboTech University, Ningbo 315100, China. He is also with Ningbo Institute, Zhejiang University, Ningbo 315100, China.
			\protect\\
			E-mail: shuzhenyu@nit.zju.edu.cn (Zhenyu Shu)
			\IEEEcompsocthanksitem Junlong Yu is with School of Software Technology, Zhejiang University, Hangzhou, PR China. Corresponding author.
			\protect\\
			E-mail: junlongyu\_paper@163.com (Junlong Yu)
			\IEEEcompsocthanksitem Kai Chao is with School of Big Data and Artificial Intelligence Management, Xi'an Jiaotong University, Xi'an, PR China.
			\IEEEcompsocthanksitem Shiqing Xin is with School of Computer Science and Technology, ShanDong University, Jinan, PR China.
			\IEEEcompsocthanksitem Ligang Liu is with Graphics \& Geometric Computing Laboratory, School of Mathematical Sciences, University of Science and Technology of China, Anhui, PR China.}
		\thanks{Manuscript received month day, year; revised month day, year.}
	}

	%
	%

\markboth{IEEE transactions on visualization and computer graphics,~Vol.~XX, No.~X, Month~Year}%
{Shell \MakeLowercase{\textit{et al.}}: Bare Demo of IEEEtran.cls for Computer Society Journals}
%



\IEEEtitleabstractindextext{%
	\begin{abstract}
		This paper presents GaussEdit, a framework for adaptive 3D scene editing guided by text and image prompts. GaussEdit leverages 3D Gaussian Splatting as its backbone for scene representation, enabling convenient Region of Interest selection and efficient editing through a three-stage process. The first stage involves initializing the 3D Gaussians to ensure high-quality edits. The second stage employs an Adaptive Global-Local Optimization strategy to balance global scene coherence and detailed local edits and a category-guided regularization technique to alleviate the Janus problem. The final stage enhances the texture of the edited objects using a sophisticated image-to-image synthesis technique, ensuring that the results are visually realistic and align closely with the given prompts. Our experimental results demonstrate that GaussEdit surpasses existing methods in editing accuracy, visual fidelity, and processing speed. By successfully embedding user-specified concepts into 3D scenes, GaussEdit is a powerful tool for detailed and user-driven 3D scene editing, offering significant improvements over traditional methods.  
	\end{abstract}
	
	}

\maketitle

\IEEEdisplaynontitleabstractindextext

%
\IEEEpeerreviewmaketitle

\begin{figure*}[t]
	\centering
	\includegraphics[width=2\columnwidth]{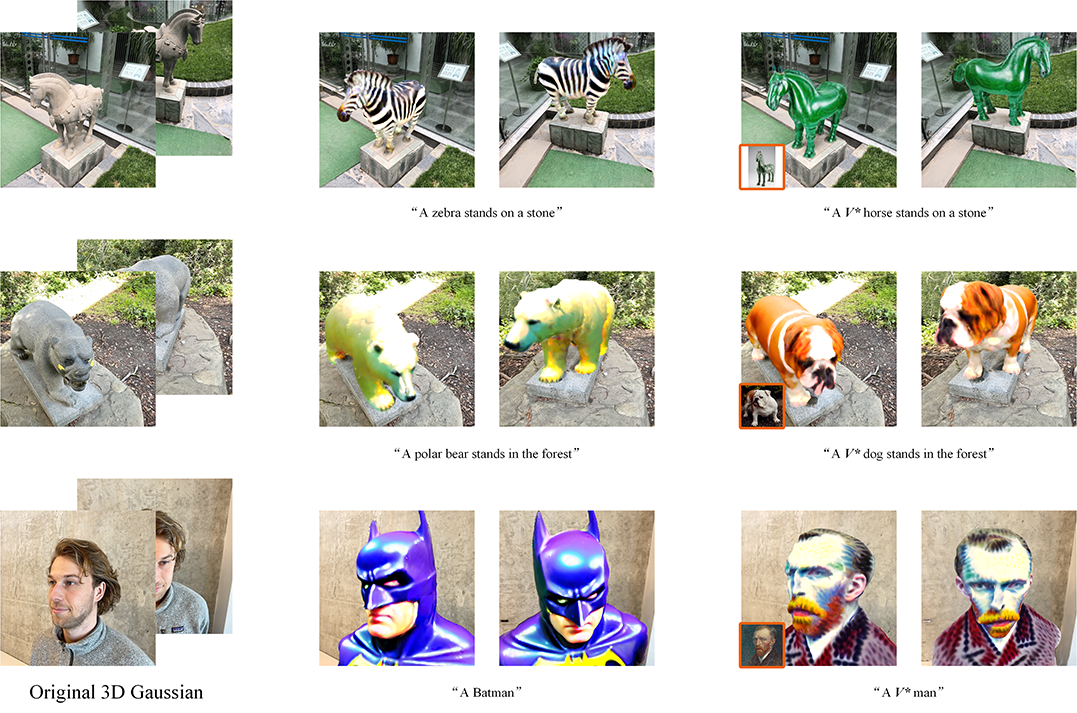}
	\caption{The edited results using our approach, GaussEdit. The first column shows the original 3D Gaussian scenes. The second column presents the scenes edited using text prompts, while the third column displays the results after editing with image prompts, utilizing reference images of specific objects. These examples highlight the effectiveness of GaussEdit in embedding user-specified concepts into 3D scenes with high fidelity and visual realism.}
	\label{fig1}
\end{figure*}

\IEEEraisesectionheading{\section{Introduction}}
\IEEEPARstart{T}{echnologies} in neural field, such as Neural Radiance Fields~(NeRF)~\cite{mildenhall2021nerf}, Neural Implicit Surfaces~(NeuS)~\cite{wang2021neus}, and 3D Gaussian Splatting~(3D-GS)~\cite{kerbl20233d}, have achieved notable progress in the fields of 3D reconstruction and novel view synthesis~\cite{li2021mine, yu2021pixelnerf, barron2021mip, cao2023scenerf, yang2022recursive, muller2022instant}. These methods can efficiently reconstruct the details of complex scenes from multi-view images while maintaining a high degree of realism by learning the geometry and texture information of scenes through deep learning models. Their impact extends to a range of downstream tasks, such as texture editing~\cite{xiang2021neutex, richardson2023texture}, shape deformation~\cite{yang2022neumesh, xu2022deforming}, scene decomposition~\cite{tang2022compressible}, and stylization~\cite{wang2023nerf}. These technologies provide unprecedented flexibility and precision in scene manipulation, significantly advancing the development of digital content creation and virtual reality applications.

With the rapid development of large-scale text-to-image~(T2I) models, generating or editing scenes solely based on text prompts~\cite{poole2022dreamfusion, haque2023instruct, sella2023vox, yi2024gaussiandreamer, zhang2024text2nerf} has unlocked new possibilities for traditional methods that rely on complex drawing and modeling~\cite{xiang2021neutex, yang2022neumesh}. However, as the saying ``a picture is worth a thousand words'' suggests, using text prompts only may fall short of capturing the intricate details and specific requirements of a user's vision. While these models are powerful, they often struggle with delivering the precision and customization needed for specialized editing tasks, leaving users with the need for additional tools or manual adjustments to achieve their desired outcome. CustomNeRF~\cite{he2023customize} is among the first to use image prompts to guide scene editing. It employs 2D image masks to identify editing regions and adopts a Local-Global Iterative strategy, achieving good results. However, it suffers from long editing times and low resolution. TIP-Editor~\cite{zhuang2024tip} focuses on using image prompts for scene editing but does not provide strong support for text-only prompts.

To address the mentioned challenge, in this paper, we propose a new approach called GaussEdit that allows users to intuitively, conveniently, and precisely edit scenes using text and image prompts. Our approach is anchored in three core designs. Firstly, we use 3D-GS as the backbone to represent the scene and extract point clouds, allowing users to precisely define the Region of Interest~(ROI) by simply setting a bounding box. We perform sampling and initialization operations on the Gaussian within the ROI to achieve faster editing results and ensure high-quality edits. Secondly, an Adaptive Global-Local Optimization strategy is adopted, which probabilistically alternates between rendering the edited objects and the entire scene to provide inputs for the T2I model. Additionally, a category-guided regularization technique is introduced, leveraging the editing category as a prompt for the multi-view consistent diffusion model. This effectively mitigates the Janus problem during the editing process. Finally, we optimize the texture of the edited objects by employing a sophisticated texture refinement technique, ensuring high-quality and realistic results in the final edited scene.

We extensively evaluated our framework, GaussEdit, across a variety of real-world and synthetic scenes, including animals, plants, humans, and outdoor settings. As demonstrated in Figures~\ref{fig1}, ~\ref{fig3}, and ~\ref{fig4}, our editing results successfully embed user-specified concepts into 3D scenes, demonstrating excellent visual effects and high-fidelity detail reproduction. These outcomes not only validate the practicality of our method but also highlight its advantages in precise control and visual fidelity, providing an effective solution for user-specified editing of complex scenes. Qualitative and quantitative comparisons confirm that GaussEdit outperforms existing methods in terms of editing accuracy, visual fidelity, and user satisfaction.

The contributions of this paper can be summarized as follows:
\begin{itemize}
	\item We propose a novel three-stage framework, GaussEdit, that uses text and image prompts to guide 3D scene editing.
	\item We present an efficient 3D Gaussian initialization method to enable fast scene editing and achieve better editing results.
	\item We design an Adaptive Global-Local Optimization strategy and a category-guided regularization technique to enhance consistency and balance optimization, thereby improving the realism of the final edited scene.
\end{itemize}

The structure of the paper is as follows. Section~\ref{rw} provides an overview of relevant research in scene editing. Section~\ref{pre} introduces the preliminaries necessary for understanding our method and sets the foundation for the subsequent sections. In Section~\ref{md}, we outline our proposed method's workflow and delve into the technical details. Section~\ref{exp} presents comprehensive experimental results and comparative analyses with baseline methods. Section~\ref{cl} concludes the paper, discusses the limitations explored, and suggests future research directions.

\section{Related Works}\label{rw}
\subsection{Text-guided image generation and editing}
The denoising diffusion probabilistic models~\cite{ho2020denoising, song2020denoising} have garnered significant attention due to their exceptional image synthesis capabilities. Subsequently, diffusion models~\cite{ramesh2022hierarchical, saharia2022photorealistic, rombach2022high} trained on large-scale image-text paired datasets have demonstrated their astonishing capability to understand complex textual prompts and generate corresponding high-quality images. These pretrained T2I models possess rich semantic understanding and high controllability, laying a solid foundation for text-guided image editing tasks~\cite{couairon2022diffedit, kawar2023imagic, hertz2022prompt, avrahami2022blended, brooks2023instructpix2pix}. Moreover, as the demand for personalized and context-specific images grows, researchers increasingly focus on tailoring image generation to align with individual styles and subject matter. 

Textual Inversion~\cite{gal2022image} and DreamBooth~\cite{ruiz2023dreambooth} are pioneering approaches to address personalization issues. Both aim to learn a specialized text token from multiple images of a single object and then generate images featuring that object. Textual Inversion~\cite{gal2022image} freezes the weights of the diffusion model UNet, whereas DreamBooth~\cite{ruiz2023dreambooth} optimizes the UNet, showing better reconstruction and generalization capabilities. Building on these two methods, Custom~Diffusion~\cite{kumari2023multi} optimizes only the cross-attention layers of the diffusion UNet and enables the generation of multiple personalized objects within a single image. CliC~\cite{safaee2023clic} focuses more locally, attempting to optimize a text token representing a visual concept that is part of an object in the image. Although these works have achieved impressive results in the 2D generation and editing field, it is non-trivial to extend them to 3D.

\subsection{Text-to-3D generation}
Thanks to the advancement of T2I models, text-to-3D generation has seen significant progress. Some efforts utilize the CLIP model to optimize meshes~\cite{michel2022text2mesh, chen2022tango, mohammad2022clip} or neural fields~\cite{jain2022zero}. A pioneering work, DreamFusion~\cite{poole2022dreamfusion}, introduced Score Distillation Sampling~(SDS), which extracts knowledge from pretrained T2I diffusion models to optimize neural fields. Subsequent works enhance this optimization process by incorporating additional refinement stages~\cite{lin2023magic3d, raj2023dreambooth3d}, integrating geometric priors~\cite{metzer2023latent}, modifying the 3D representation~\cite{chen2023fantasia3d}, or proposing more effective variants of SDS~\cite{wang2024prolificdreamer} to improve the quality of the generated outcomes. To alleviate the Janus problem, Perp-neg~\cite{armandpour2023re} uses perpendicular denoising to preserve main concepts while eliminating unwanted attributes. Gaussiandreamer~\cite{yi2024gaussiandreamer} leverages 3D diffusion to generate point cloud priors, which are then optimized using T2I diffusion to ensure 3D consistency. 

However, due to the difficulty of perfectly aligning existing scenes with text, all these methods are geared more towards generating new scenes rather than editing existing 3D scenes.

\subsection{Text-to-3D Editing}
Neural Fields, which use deep neural networks to represent 3D scenes, have garnered significant interest. In response to their growing utility, numerous works have begun to explore the editing of Neural Fields. Some research focuses on interactive editing, either by manipulating appearance latent~\cite{liu2021editing, park2021hypernerf}, interacting with proxy representations~\cite{yang2022neumesh, yuan2022nerf}, or using segmentation areas and masks~\cite{kobayashi2022decomposing, kuang2023palettenerf, mirzaei2022laterf}. Other studies employ text-driven approaches to edit Neural Fields. Instruct-Nerf2Nerf~\cite{haque2023instruct} edits rendered images based on text prompts and then uses these images to update the NeRF. In contrast, VoX-E~\cite{sella2023vox} and DreamEditor~\cite{zhuang2023dreameditor} use explicit 3D representations—voxels and meshes, respectively—and extract 2D cross-attention maps from the diffusion model UNet to identify target regions and apply SDS loss to perform local edit. TIP-Editor~\cite{zhuang2024tip} employs a stepwise 2D personalization strategy to fine-tune the T2I diffusion model. It then calculates both global and local SDS losses to edit the scene effectively. ED-NeRF~\cite{park2023ed} adopts a NeRF backbone similar to Latent-NeRF~\cite{metzer2023latent}, embedding 3D scenes into the latent space of the Latent Diffusion Model~(LDM)~\cite{rombach2022high} to accelerate training speeds and uses Delta Denoising Score~(DDS)~\cite{hertz2023delta} loss to enhance editing effects. In addition, Instruct 4D-to-4D~\cite{mou2024instruct} treats 4D scenes as pseudo-3D scenes, enabling video editing techniques and 3D scene editing methods to edit 4D scenes effectively.

The work most similar to ours is CustomNeRF~\cite{he2023customize}, which suffers from limitations such as low resolution and long editing times. In contrast, our approach offers higher resolution, faster editing speeds, and improved 3D consistency, providing a more efficient and accurate solution for 3D scene editing.

\section{Preliminaries}\label{pre}
\subsection{Score Distillation Sampling}
Score Distillation Sampling, proposed by DreamFusion~\cite{poole2022dreamfusion}, is a distillation technique used for pretrained T2I diffusion models, effectively applied in 3D generation. It leverages the T2I diffusion model as guidance to optimize the 3D representation. 

Specifically, it uses NeRF as the 3D representation, optimizing its parameters $\theta$. First, an image $x = g(\theta)$ is rendered from a random viewpoint, where $g(\cdot)$ denotes the rendering method. To ensure that the rendered image $x$ is similar to the samples obtained from the diffusion model $\phi$, a score estimation function $\hat{\epsilon}_{\phi}(z_t; y, t)$ is introduced. Given a noisy image $z_t$, a noise level $t$, and a text prompt $y$, predicts the sampled noise $\hat{\epsilon}_{\phi}$. By comparing the predicted noise $\hat{\epsilon}_{\phi}$ with the Gaussian noise $\epsilon$ added to the rendered image $x$, the score estimation function provides a direction for updating the parameters $\theta$. The gradient computation is formulated as follows:
\begin{equation}
	\nabla_{\theta} \mathcal{L}_{SDS} (\phi, x = g(\theta)) = \mathbb{E}_{t, \epsilon} \left[ w(t) ( \hat{\epsilon}_{\phi} (z_t; y, t) - \epsilon ) \frac{\partial x}{\partial \theta} \right],
	\label{eq1}
\end{equation}
where $w(t)$ is a weighting function that depends on the time step $t$.

\subsection{3D Gaussian Splatting}
Compared to NeRF~\cite{mildenhall2021nerf}, the recently proposed 3D Gaussian Splatting~\cite{kerbl20233d} has garnered significant attention for novel-view synthesis due to its high rendering quality and real-time performance. Specifically, 3D-GS represents a scene through a set of anisotropic Gaussians, each containing five optimizable attributes: center position $\mu \in \mathbb{R}^3$, scaling factor $s \in \mathbb{R}^3$, rotation quaternion $q \in \mathbb{R}^4$, color $c \in \mathbb{R}^3$, and opacity $\alpha \in \mathbb{R}^1$. The location of  3D Gaussian can be expressed as follows:
\begin{equation}
	G(x) = e^{-\frac{1}{2} x^T \Sigma^{-1} x}, \Sigma = qss^Tq^T,
	\label{eq2}
\end{equation}
where $x$ represents the distance between $\mu$ and the query point. To compute the color of each pixel, 3D-GS uses a typical neural point-based rendering method. Specifically, a ray $r$ is cast from the camera center, and the color and density at the intersections of the ray with the 3D-GS are computed. The rendering process is as follows:
\begin{equation}
	C(r) = \sum_{i \in \mathcal{N}} c_i \sigma_i \prod_{j=1}^{i-1} (1 - \sigma_j), \quad \sigma_i = \alpha_i G(x_i),
	\label{eq3}
\end{equation}
where $N$ denotes the number of sample points along the ray $r$, and $c_i$, $\alpha_i$, and $x_i$ represent the color, opacity, and distance to the center point of the $i$-th Gaussian, respectively.

\section{Methodology}\label{md} 

\begin{figure*}[t]
	\centering
	\includegraphics[width=1.88\columnwidth]{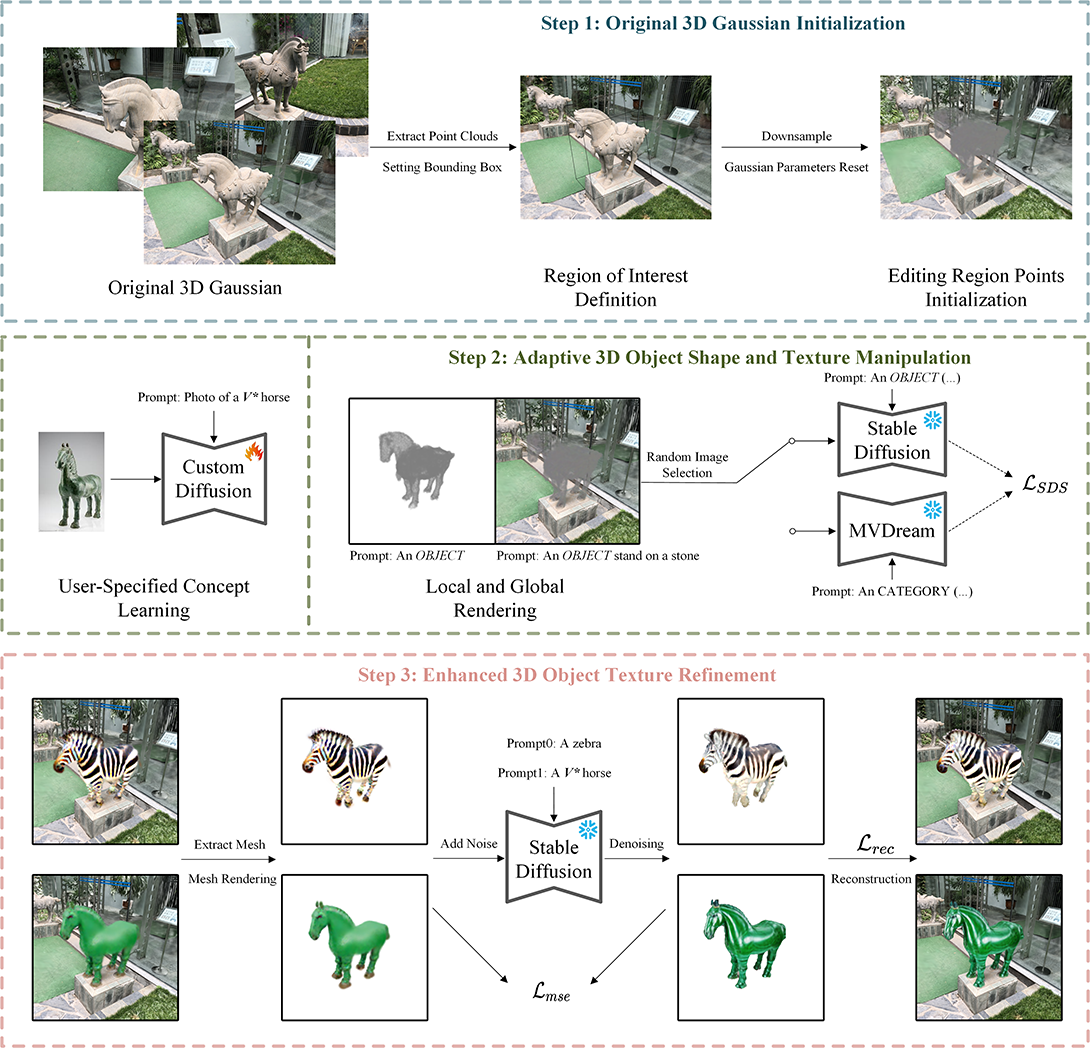}
	\caption{Our GaussEdit enables editing a 3D scene using textual descriptions or reference images through a streamlined pipeline. Initially, we extract point clouds and set bounding boxes to define the Region of Interest. Subsequently, we downsample and reset the Gaussian parameters of the editing region's points. For the reference image, we utilize the Custom Diffusion method for user-specified concept learning and generate a special token ``\textbf{\textit{V*}}''. Subsequently, we apply an Adaptive Global-Local Optimization strategy, selecting the input image and its corresponding prompt. This is followed by a category-guided regularization technique, where parameters are alternately optimized using Stable Diffusion and MVDream. When the prompt is processed by MVDream, the placeholder ``OBJECT'' in the prompt is replaced with its corresponding ``CATEGORY''. The preliminary editing results often contain noise; thus, we refine the object's texture using image-to-image synthesis. By incorporating the T2I diffusion model, we add noise to the rendered images and then perform denoising to ensure the final images align with the editing requirements.}
	\label{fig2}
\end{figure*}

Given a set of posed images of the target scene, our goal is to guide the editing of the scene using text or image prompts. We use 3D-GS to represent the scene because of its efficient handling of point cloud data, allowing convenient selection of editing regions using bounding boxes and other benefits such as high rendering quality and speed. As illustrated in Figure~\ref{fig2}, our method consists of three steps. First, we export the reconstructed scene as a point cloud, select the objects to be edited using bounding boxes, and perform sampling and initialization of the Gaussians in these regions. Second, we adopt an Adaptive Global-Local Optimization strategy, probabilistically feeding rendered images of the editing parts or the entire scene and the corresponding prompts into T2I models. We then employ a category-guided regularization technique, alternately utilizing Stable Diffusion and MVDream~\cite{shi2023mvdream} to optimize the Gaussian parameters. During this process, the original editing terms in the prompts are replaced with category-specific descriptors when using MVDream. For image prompts, we first encode the image into a special token ``\textbf{\textit{V*}}'' using the method proposed by Custom Diffusion~\cite{kumari2023multi}. Finally, after multiple iterations to achieve an initial shape and texture, we use an image-to-image synthesis approach to enhance the texture.

\subsection{Original 3D Gaussian Initialization}\label{m1}
The initialization of 3D-GS is a crucial step in our proposed method to ensure accurate and efficient editing of the scene. We start by exporting the reconstructed scene as a point cloud, which provides a detailed representation of the scene's geometry.

Assuming the original set of Gaussians is $\mathcal{G}$, where $\mathcal{G}i = {\mu_i, s_i, q_i, c_i, \alpha_i}$ represents the parameters for the $i$-th Gaussian. The first step involves defining the ROI by setting a bounding box around the desired area for editing. Once the bounding box is set, we can divide the scene into two distinct parts: the Gaussians inside the bounding box, designated as the set of editable Gaussians $\mathcal{G}{in}$, and the Gaussians outside the bounding box, forming the set of non-editable Gaussians $\mathcal{G}_n$.

Next, we perform farthest point sampling on $\mathcal{G}_{in}$ to ensure a uniform distribution of points. This step avoids clustering and ensures that the entire region is adequately covered. The sampled points are then initialized with specific attributes: random colors, unit scaling, and no rotation. Additionally, the opacity $\alpha$ is set to a small value, $\log(0.1/0.9)$. The resulting set of editable Gaussians, now initialized with these attributes, is denoted as $\mathcal{G}_e$.

Finally, by concatenating the non-editable and editable sets, we obtain the initialized Gaussians for the entire scene:
\begin{equation}
	\mathcal{G}_{all} = \mathcal{G}_n \oplus \mathcal{G}_e,
	\label{eq4}
\end{equation}
where $\oplus$ denotes the concatenation operation.

\subsection{Adaptive 3D Object Shape and Texture Manipulation}\label{m2}
In this step, we focus on manipulating the shape and texture of the 3D object guided by text or image prompts. To ensure uniformity and coherence, we employ an efficient User-Specified Concept Learning method, Custom Diffusion~\cite{kumari2023multi}, to align the reference image with the text ``\textbf{\textit{V*}}''.

An intuitive method to edit the origin scene involves directly optimizing $\mathcal{G}_e$ using the SDS loss calculated by a pre-trained T2I diffusion model, as done in GaussianDreamer~\cite{yi2024gaussiandreamer}. The optimized $\mathcal{G}_e$ is then combined with $\mathcal{G}_n$ to produce the final result. However, due to the object's position, size, and constraints within the scene, directly optimizing $\mathcal{G}_e$ often leads to suboptimal outcomes that can result in distorted shapes and textures that do not integrate well with the overall scene. To address this, we propose a more nuanced approach that considers the spatial and contextual factors affecting the object. This ensures a more coherent and aesthetically pleasing integration of the 3D object into the desired environment.

By considering these factors, in each iteration, we render both local and global images and select the appropriate one based on a probability threshold:
\begin{equation}
	I =
	\begin{cases}
		g(\mathcal{G}_e), & \text{if}\; r < p \\
		g(\mathcal{G}_{all}), & \text{if}\; r \ge p
	\end{cases}.
	\label{eq5}
\end{equation}
Here, $g$ represents the splatting rendering method as defined in Equation~\ref{eq3}, and $r$ is a random number between $[0,1]$. Once the image $I$ is selected, we can calculate the SDS loss as follows:
\begin{equation}
	\nabla_{\mathcal{G}_e} \mathcal{L}_{SDS} (\phi, I) = \mathbb{E}_{t, \epsilon} \left[ w(t) ( \hat{\epsilon}_{\phi} (z_t; y, t) - \epsilon ) \frac{\partial I}{\partial \mathcal{G}_e} \right],
	\label{eq6}
\end{equation}
where $y$ is the prompt corresponding to $I$. For instance, in Step 2 of Figure~\ref{fig2}, if $r<p$, the prompt is ``A \textbf{\textit{V*}} horse''; otherwise, it is ``A \textbf{\textit{V*}} horse stands on a stone.''

In our experiments, we observed that using Stable Diffusion often leads to a significant Janus problem. To mitigate this issue, we propose a category-guided regularization technique. This approach incorporates an additional multi-view consistent T2I model, MVDream, and alternates between the two T2I models for parameter optimization. When utilizing MVDream in a given iteration, we replace the editing term in the prompt with its corresponding category term. For example, the prompt ``\textbf{\textit{V*}} horse'' is replaced with ``horse''. 

\subsection{Enhanced 3D Object Texture Refinement}\label{m3}
As illustrated in Step 3 of Figure~\ref{fig2}, optimizing using SDS loss can introduce artifacts, such as colored noise on the zebra's body, which detracts from the realism of the edited scene. To address this issue, we aim to enhance the details given the coarse texture.

Following the approach of SDEdit~\cite{meng2021sdedit}, which utilizes stochastic differential equations to reverse-solve and generate images, we can achieve a natural balance between realism and fidelity~(input relevance). First, we extract the edited part of the scene and export it as a mesh. Then, the mesh is rendered from arbitrary viewpoints to obtain the image $I_r$. Next, we add random noise at timestep $t$ to $I_r$ and perform a multi-step denoising process $f_{\phi}(\cdot)$ using the T2I diffusion model to obtain the refined image $I_d$:
\begin{equation}
	I_d = f_{\phi}(z_t; y, t).
\end{equation}
This denoising process helps to remove the artifacts introduced by the initial optimization, smoothing out any irregularities and enhancing the overall texture quality. The resulting image $I_d$ is then compared to the original rendered image $I_r$ to ensure the refinement maintains the essential details and structure. To achieve this, we apply a Mean Squared Error loss $\mathcal{L}_{mse}$ to optimize the texture of the mesh:
\begin{equation}
	\mathcal{L}_{mse} = \left\| I_d -I_r \right\|^2.
\end{equation}
After multiple iterations, we perform a reconstruction of the edited part, using a reconstruction loss $\mathcal{L}_{rec}$ to optimize $\mathcal{G}_e$ :
\begin{equation}
	\mathcal{L}_{rec} = (1 - \lambda) \left | I - I_r \right | + \lambda \mathcal{L}_{D-SSIM},
\end{equation}
where $\lambda$ is a weighting factor that balances the components of the reconstruction loss, and $\mathcal{L}_{D-SSIM}$ is the structural similarity index measure.

By integrating this approach, we enhance the texture details of the 3D object while maintaining coherence with the original scene. This dual focus on pixel-level accuracy and structural similarity ensures that the refined textures are realistic and contextually appropriate. As shown in Figure~\ref{fig2}, this process ensures high-quality textures and a more realistic final rendering. The integration of these steps allows for detailed and coherent enhancements, making the edited scene visually appealing to the original context.

\begin{figure*}
	\centering
	\includegraphics[width=1.9\columnwidth]{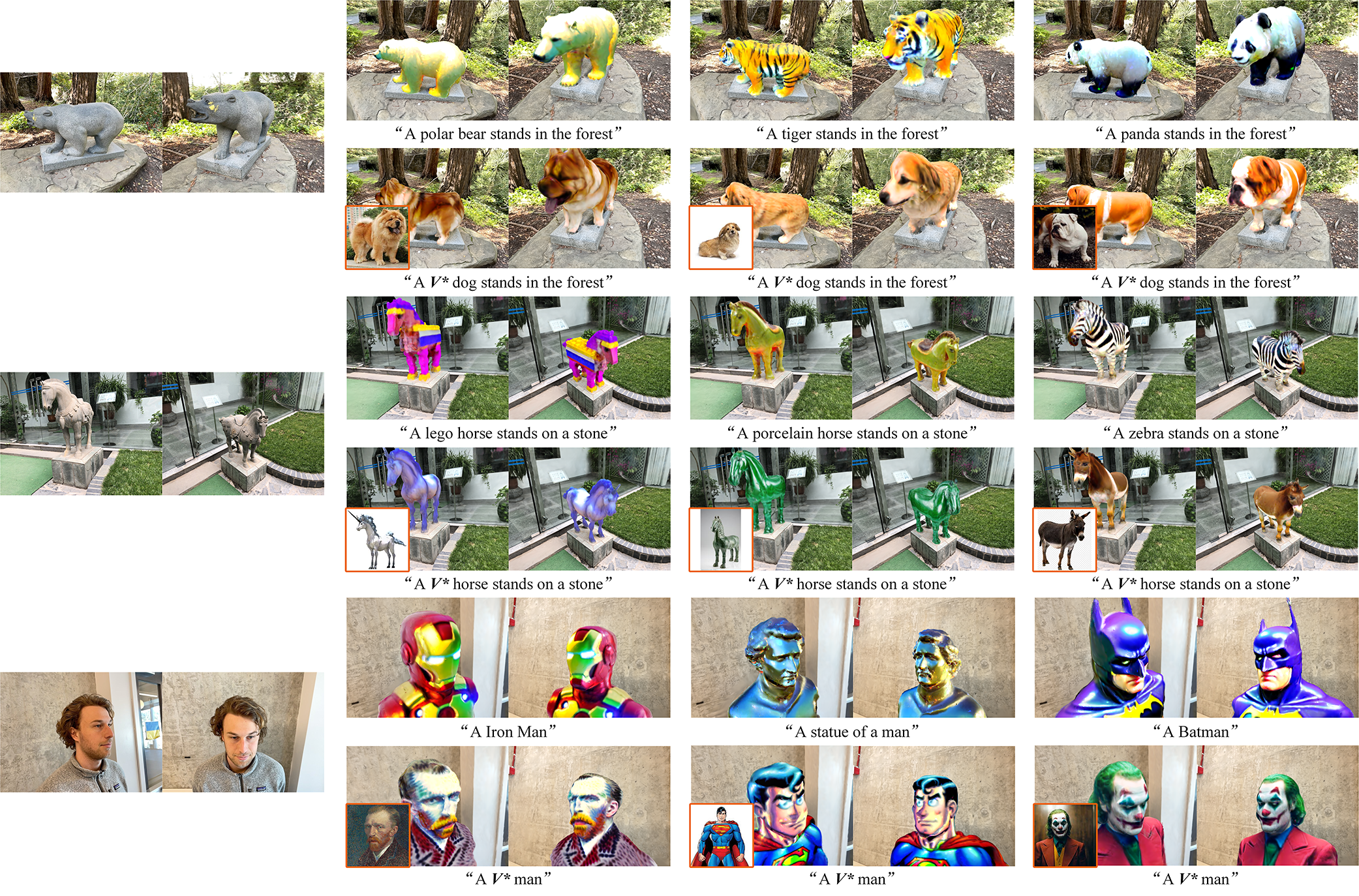}
	\caption{Visual results of our proposed GaussEdit. The first column shows the original scene, while the next three display the edited results. Each scene is guided by text and image prompts, with the reference image placed in the bottom left corner of the edited results.}
	\label{fig3}
\end{figure*}

\begin{figure*}
	\centering
	\includegraphics[width=1.9\columnwidth]{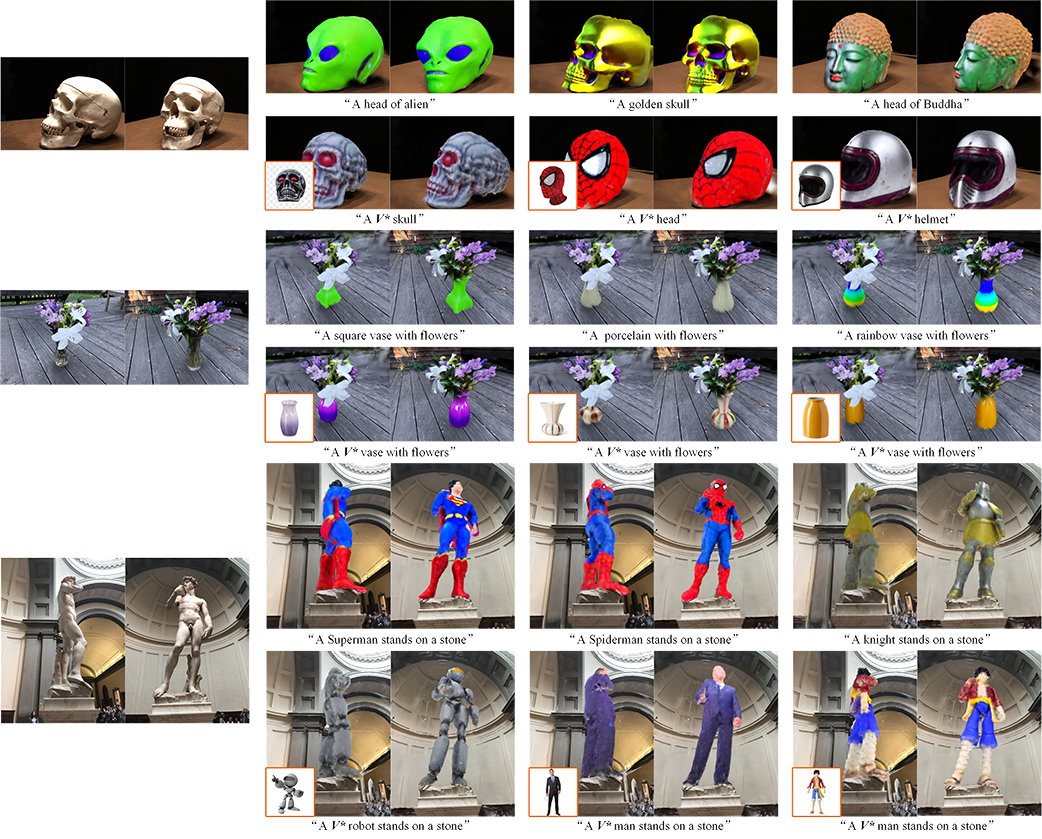}
	\caption{Visual results of our proposed GaussEdit. The first column shows the original scene, while the next three display the edited results. Each scene is guided by text and image prompts, with the reference image placed in the bottom left corner of the edited results.}
	\label{fig4}
\end{figure*}

\begin{figure*}
	\centering
	\includegraphics[width=1.9\columnwidth]{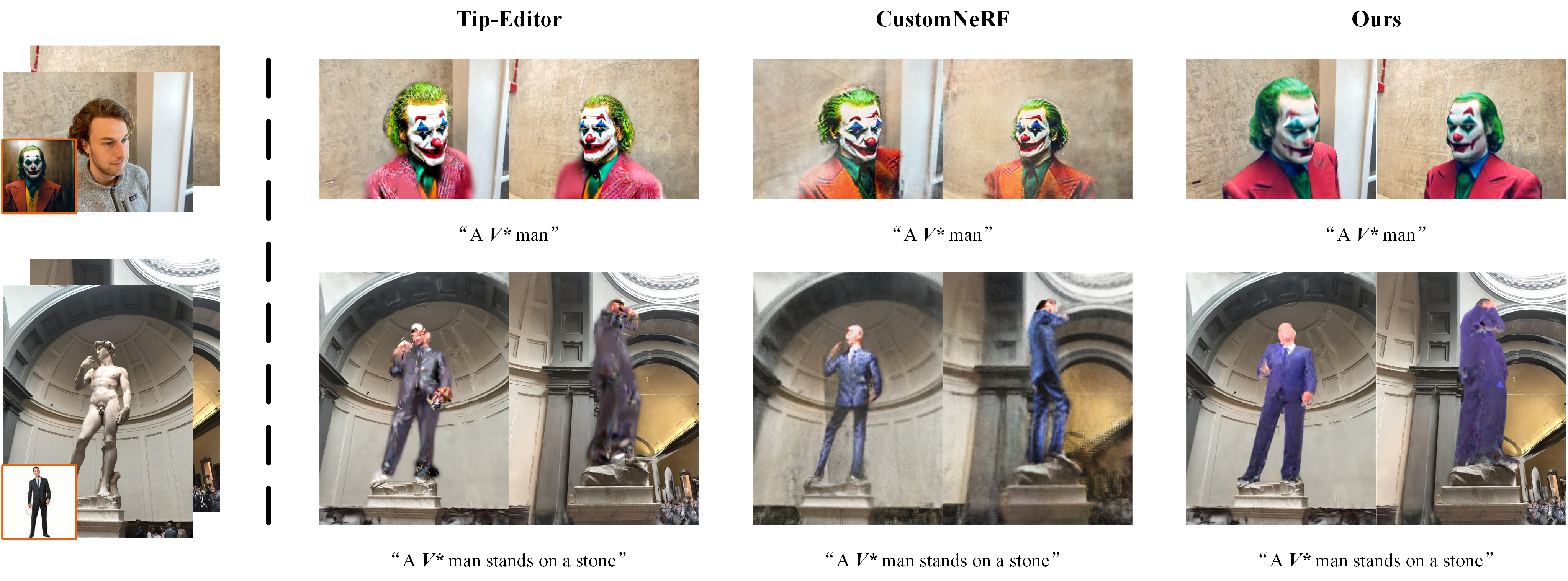}
	\caption{Qualitative comparison on the image-driven editing setting.}
	\label{fig5}
\end{figure*}

\begin{figure*}
	\centering
	\includegraphics[width=1.9\columnwidth]{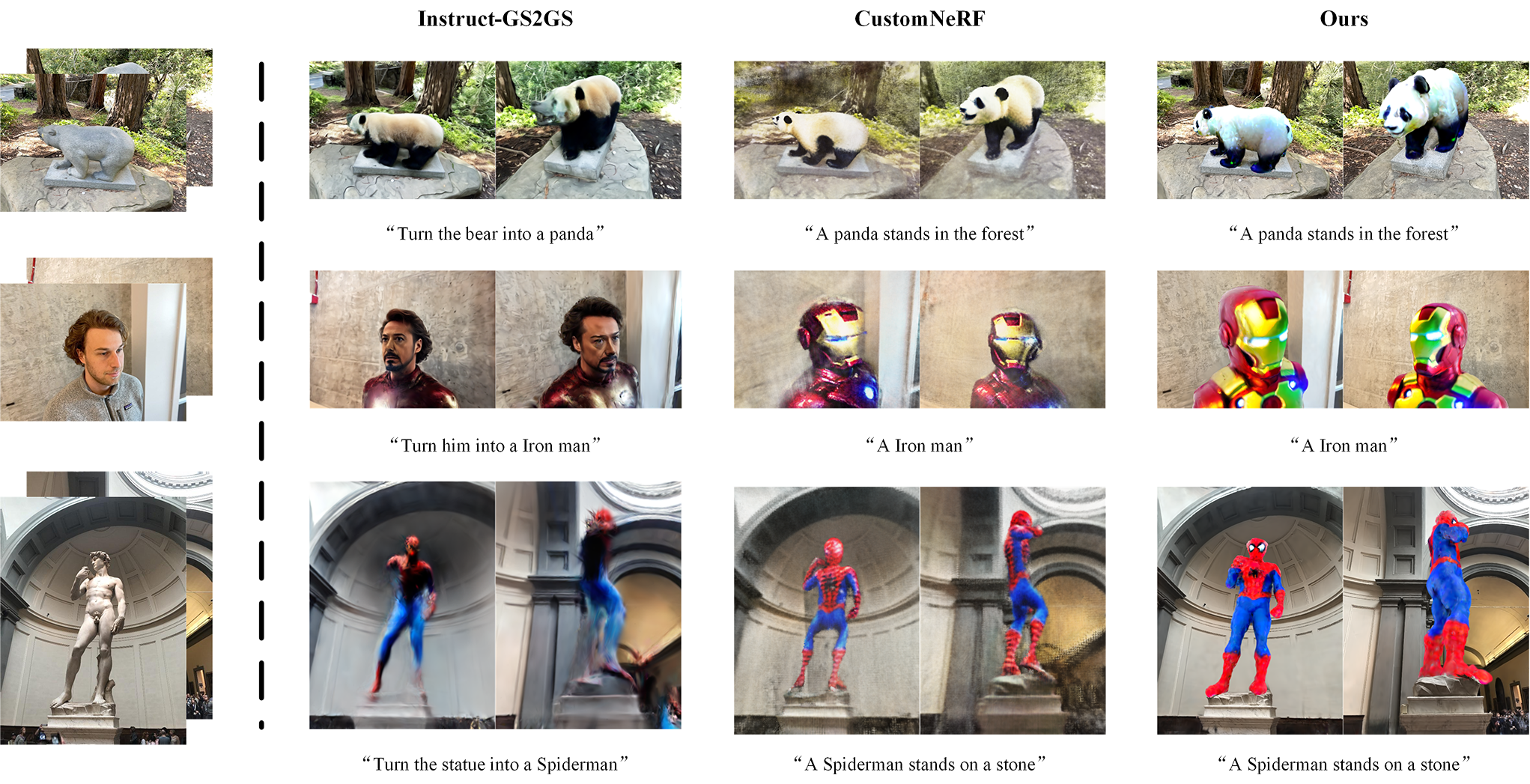}
	\caption{Qualitative comparison on the text-driven editing setting.}
	\label{fig6}
\end{figure*}

\section{Experiments}\label{exp}
\subsection{Experimental Setup}
\textbf{Implementation Details.} Our method is implemented using PyTorch. For original scene reconstruction, we use the default parameters as in~\cite{kerbl20233d}. During the Original 3D Gaussian Initialization step, we set the number of farthest point sampling to 50,000. In the Adaptive 3D Object Shape and Texture Manipulation step, the number of iterations is set between 1,200 and 1,600, taking approximately 8-12 minutes. The learning rates are set as follows: center position $\mu$ at 1e-3, scaling factor $s$ at 5e-3, rotation quaternion $q$ at 1e-3, and opacity $\alpha$ at 5e-2. The color $c$ of 3D-GS is represented by spherical harmonics coefficients of degree 0, with a learning rate of 5e-2. 

For the T2I diffusion models, we use Stable~Diffusion~v1.5 and MVDream~\cite{shi2023mvdream}. For image-driven editing, we follow the Custom~Diffusion training process to conduct user-specified concept learning with 250 steps. The timestamps we use are uniformly sampled from 0.02 to 0.98 before 600 iterations and change to 0.02 to 0.55 after 600 iterations. 

In the Enhanced 3D Object Texture Refinement, we set the number of iterations from 0 to 200 based on the results of the previous step, taking approximately 0-5 minutes. In all of our experiments, the rendering resolution is set to 512x512 with a batch size of 4. All experiments are conducted on a single RTX 3090.

\textbf{Dataset.} To evaluate the effectiveness of our GaussEdit method, we selected six representative scenes with varying background complexities, including animals, plants, human, and outdoor settings, as described in ~\cite{yao2020blendedmvs, mildenhall2021nerf, haque2023instruct, he2023customize}. For each scene, we used COLMAP~\cite{schonberger2016structure} to estimate camera poses and reconstruct the original 3D-GS. We used six prompts to guide the edits. For image-driven editing, we downloaded a reference image from the Internet.

\textbf{Baselines.} For editing methods guided solely by text prompts, we compare GaussEdit with Instruct-GS2GS~\cite{igs2gs}, which iteratively updates Gaussian by modifying the image dataset using a T2I diffusion model. Instruct-GS2GS is the 3D-GS implementation of Instruct-NeRF2NeRF~\cite{haque2023instruct} and demonstrates superior editing speed and quality than Instruct-NeRF2NeRF. We also compare to CustomNeRF~\cite{he2023customize}, which uses 2D image masks and a Local-Global Iterative strategy for scene editing.

For image-driven approaches, we compare our method with TIP-Editor~\cite{zhuang2024tip}, which employs a stepwise 2D personalization strategy to fine-tune the T2I diffusion model, and CustomNeRF~\cite{he2023customize}.

\textbf{Evaluation Criteria.} Following the evaluation methods in~\cite{haque2023instruct, he2023customize, zhuang2024tip, zhuang2023dreameditor}, we evaluate GaussEdit using two criteria CLIP Text-Image directional similarity and DINO similarity.

For text-driven editing, we use CLIP Text-Image directional similarity, which assesses the alignment between two images (original image $i_o$ and edited image $i_e$) and two text prompts (original text description $p_o$ and editing text prompt $p_e$). This reflects whether the direction of image editing is consistent with the direction of text change. The mathematical formula is as follows:
\begin{subequations}
	\begin{align}
		\bigtriangleup P &= Text(p_e) - Text(p_o), \\
		\bigtriangleup I &= Image(i_e) - Image(i_o), \\
		CLIP_{dir} &= 1 - \frac{\bigtriangleup P \cdot \bigtriangleup I}{|\bigtriangleup P| |\bigtriangleup I|},
	\end{align}
	\label{eq9}
\end{subequations}
where $Text(\cdot)$ and $Image(\cdot)$ represent the text and image encoders of CLIP, respectively. We obtain the original text description $p_o$ using BLIP-2~\cite{li2023blip} and modify it to get $p_e$.

For image-driven editing, the evaluation method is DINO similarity. To assess whether the edited scene aligns with the reference image, we can determine the consistency by calculating the average DINO similarity between the edited images from various perspectives and the reference image. The formula is as follows:

\begin{equation}
	Sim = \frac{Enc(i_e) \cdot Enc(i_r)}{|Enc(i_e)| |Enc(i_r)|},
	DINO_{sim} = \frac{Sim+1}{2},
	\label{eq10}
\end{equation}
where $Enc(\cdot)$ is the image encoder of DINOv2~\cite{oquab2023dinov2} and $i_r$ is the reference image.

\subsection{Qualitative Results}
\textbf{Editing guided by image prompts.} Figure~\ref{fig5} presents the results of our scene editing using reference images as prompts. As shown in the second column, the final output of the Tip-Editor method still exhibits some noise. We think this is due to the refining step in Tip-Editor, where optimized versions of multiple rendered images are used as ground truth. This step employs reconstruction loss to optimize the scene, potentially causing multi-view inconsistencies that fail to eliminate noise effectively. In contrast, our refining step adopts an iterative dataset update approach similar to Instruct-NeRF2NeRF. This method optimizes based on the current iteration's scene, resulting in a smoother final output. Compared to the other two methods, our approach demonstrates superior realism and achieves more robust separation between the foreground and background. However, the one limitation shared by all three methods is the inability to learn the concept from the reference image thoroughly. We attribute this to the inherent limitations of relying on fine-tuning T2I models and optimizing the scene using SDS loss. This approach may struggle to capture the fine-grained details or complex semantics of the reference image, impacting the final editing results.

\textbf{Editing guided by text prompts.} Figure~\ref{fig6} presents the results of our scene editing using text prompts. Compared to Instruct-GS2GS, our method performs better in terms of structural integrity and semantic changes in the edited results. For instance, when transforming a bear into a panda, the results from the Instruct-GS2GS method still retain remnants of the original bear, whereas, in the scene edited to Iron Man, there is a noticeable discrepancy between the edited result and the expected outcome. This improvement stems from our approach of separating the foreground and background and using high-resolution images for training. Additionally, our method demonstrates better consistency. For instance, in the scene edited to Iron Man, CustomNeRF produces inconsistent results in the eye region. This issue is mitigated in our method due to the use of a category-guided regularization technique, which incorporates a multi-view consistent T2I diffusion model. Overall, our method delivers more ideal editing results with enhanced realism and consistency.

\begin{table}[h]
	\centering
	\caption{Quantitative comparison on the image-driven editing setting. DINO$_{sim}$ is the DINO similarity.}
	\label{tab1}
	\renewcommand\tabcolsep{4pt}
	\small
	\begin{tabular}{lccc}
		\toprule
		Method   & DINO$_{sim} \uparrow$  & Time(Minutes) & Vote$_{Percentage}$ \\
		\midrule
		Tip-Editor   & 0.712 & 20  & 37.6\% \\
		CustomNeRF   & 0.673 & 30  & 21.2\% \\
		Ours & \textbf{0.719} & \textbf{10-15} & \textbf{41.2\%} \\
		\bottomrule
	\end{tabular}
\end{table}

\begin{table}[h]
	\centering
	\caption{Quantitative comparison on the text-driven editing setting. CLIP$_{dir}$ denotes the CLIP Text-Image directional similarity.}
	\label{tab2}
	\renewcommand\tabcolsep{4pt}
	\small
	\begin{tabular}{lccc}
		\toprule
		Method  & CLIP$_{dir} \uparrow$   & Time(Minutes) & Vote$_{Percentage}$ \\
		\midrule
		Instruct-GS2GS   & 23.55 & 30  & 15.7\%\\
		CustomNeRF   & 26.82 &  30 & 35.9\%\\
		Ours & \textbf{29.13} & \textbf{10-15} & \textbf{48.3\%} \\
		\bottomrule
	\end{tabular}
\end{table}

\subsection{Quantitative results}
Table~\ref{tab1} and Table~\ref{tab2} present the quantitative comparison results of our method against other approaches for image-driven and text-driven editing tasks, respectively. We distributed 50 questionnaires to evaluate the realism of the editing results, the degree of background distortion, image resolution, and the alignment with the target requirements, and then we calculated the average vote rate. From the results in both tables, it is evident that our method achieved the highest approval ratings. Moreover, our method outperformed the others in  DINO similarity and CLIP Text-Image directional similarity, indicating that our editing results align more closely with both text and image prompts. Additionally, our approach demonstrates superior time efficiency.

\subsection{Ablation study}
\begin{figure}[h]
	\centering
	\includegraphics[width=0.9\columnwidth]{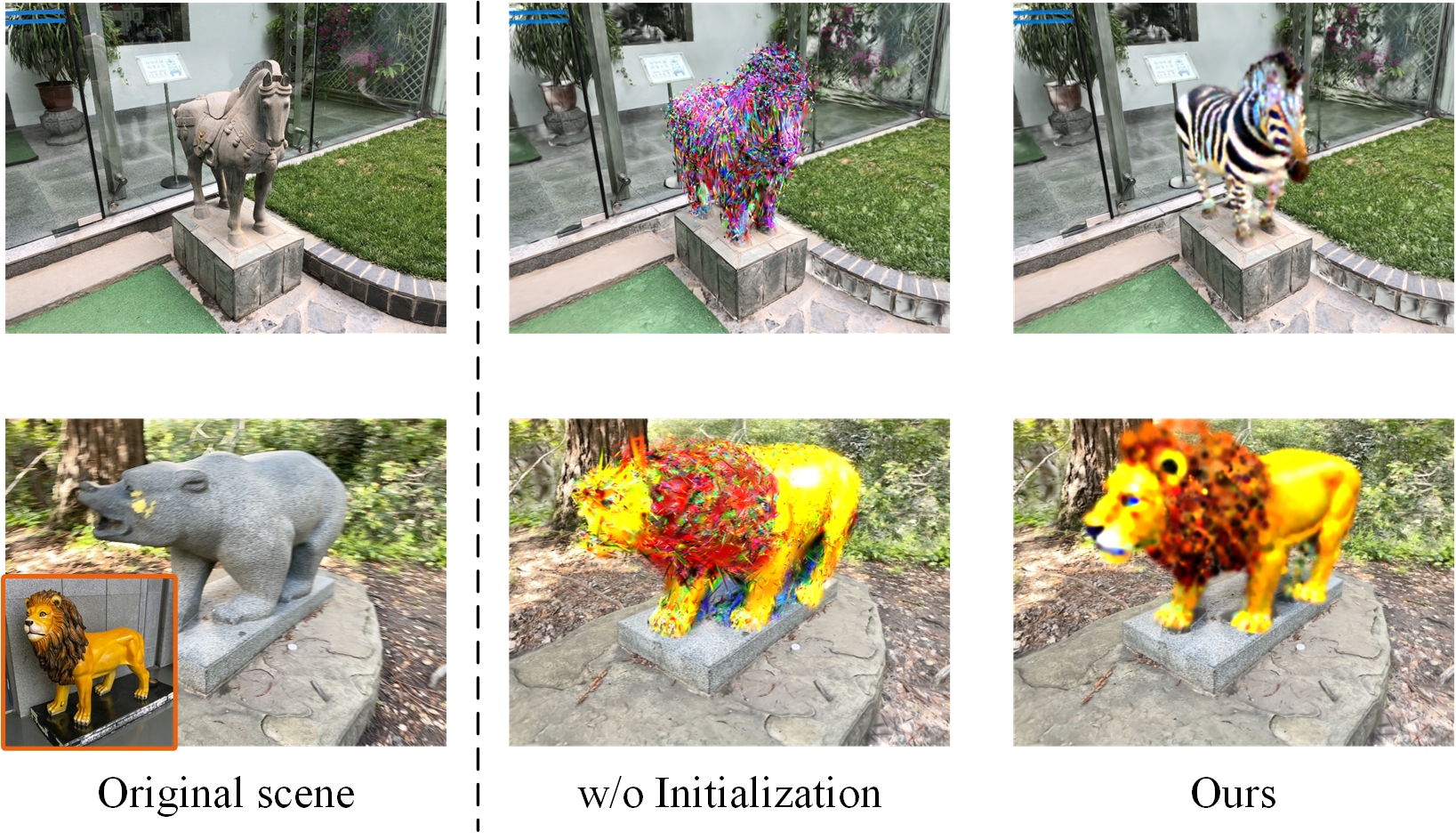}
	\caption{Ablation study of Gaussian Initialization. We conducted an ablation study to evaluate the impact of Gaussian Initialization in both text-driven and image-driven scenarios. The term ``w/o Initialization'' indicates results without applying Gaussian Initialization.}
	\label{fig7}
\end{figure}
\textbf{Effectiveness of Gaussian Initialization.} Figure~\ref{fig7} demonstrates the effectiveness of Gaussian Initialization (Section~\ref{m1}) through an ablation study. Under the ``w/o Initialization'' scenario, the Gaussian parameters of the scene are used as they are after reconstruction, without any additional preparation or parameter adjustment tailored for editing. As depicted in the figure, applying the Gaussian Initialization step while keeping all other parameters constant results in superior outcomes. For instance, the shapes and textures of the zebra and lion are more accurately generated in the third column. This enhancement can be attributed to the smoother starting conditions provided by Gaussian Initialization, which aids the optimization process in converging more efficiently. The results shown in the figure have not been subjected to the Enhanced 3D Object Texture Refinement step (Section~\ref{m3}). This omission is intentional because the quality of the second stage's editing results directly affects the outcome of the third stage. Therefore, we present the editing results from the second stage to provide a clear representation.

\begin{figure}[h]
	\centering
	\includegraphics[width=0.9\columnwidth]{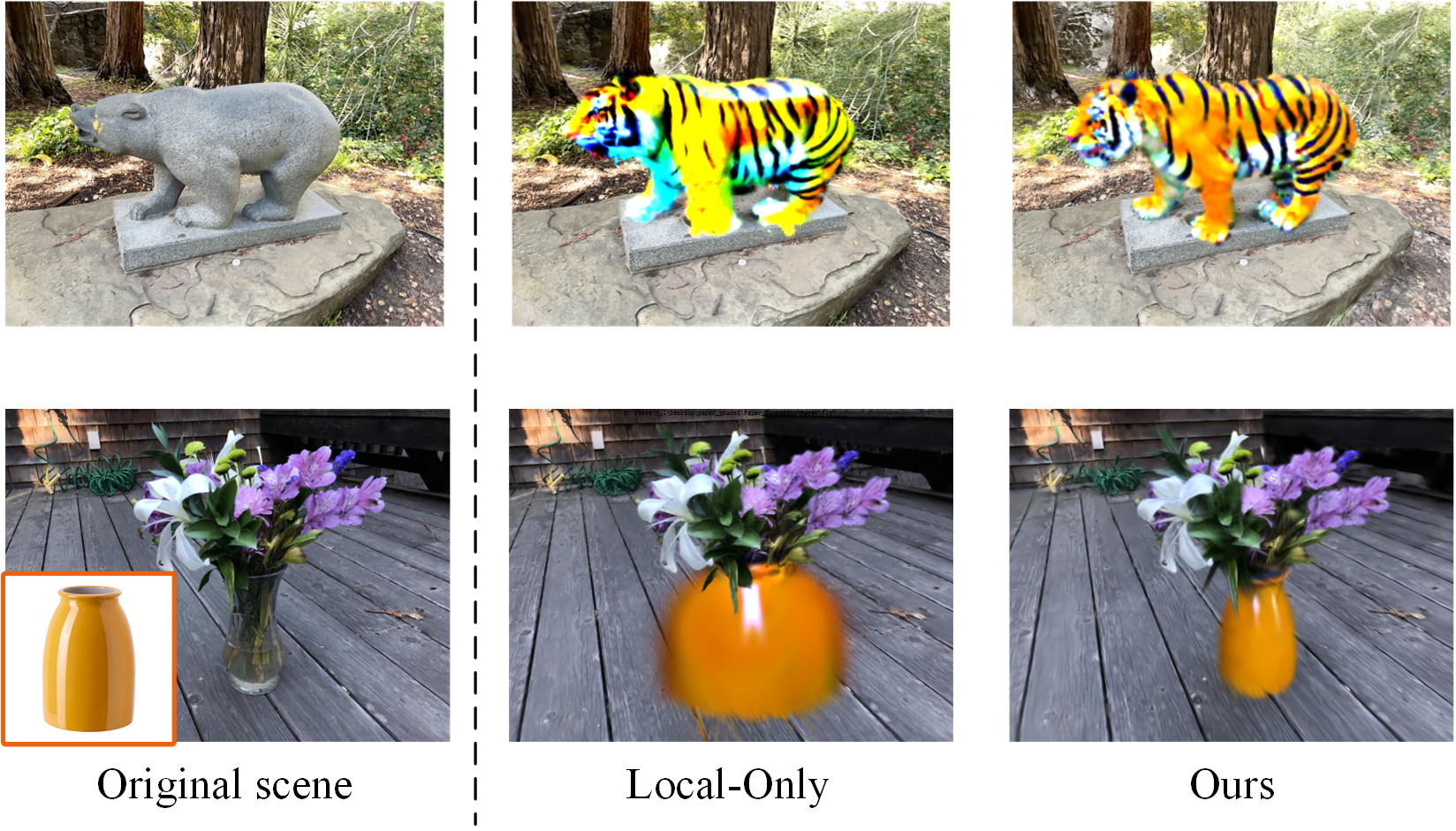}
	\caption{Ablation study of Adaptive Global-Local Optimization strategy. ``Local-Only'' indicates results where the scene is edited using only the local SDS loss.}
	\label{fig8}
\end{figure}

\textbf{Effectiveness of Adaptive Global-Local Optimization strategy.} Figure~\ref{fig8} presents the ablation results of employing the Adaptive Global-Local Optimization strategy. When only local settings are used, editing without taking the global context into account can lead to a mismatch between the content of the foreground and background. For instance, in the second column, the relative positions and sizes of the tiger and the vase are incongruent with the overall scene. This issue is effectively resolved by incorporating global background information, as shown in the third column, ensuring that the edited results are consistent with the prompt and harmonious within the scene.

\begin{figure}[h]
	\centering
	\includegraphics[width=0.9\columnwidth]{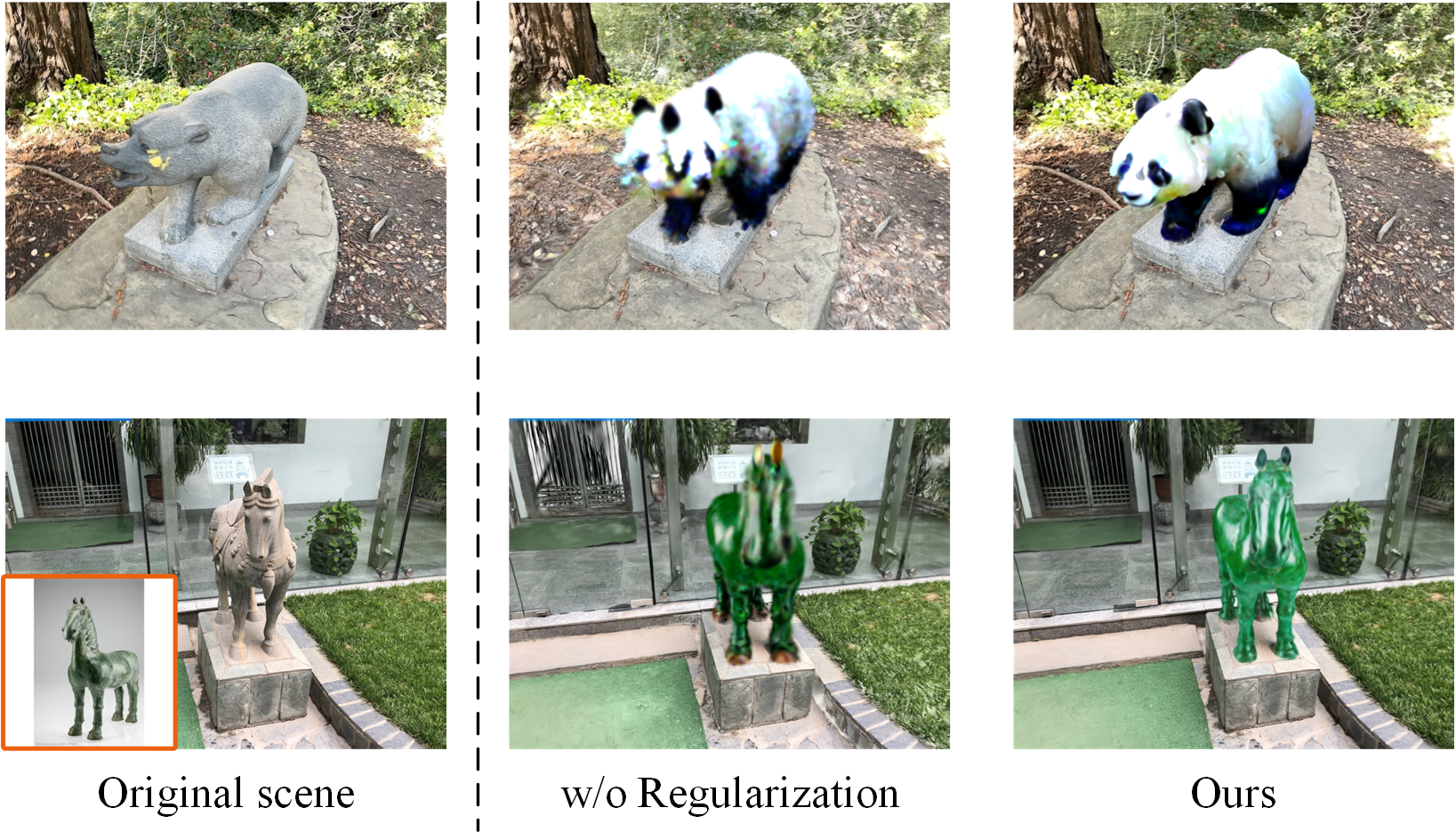}
	\caption{Ablation study of category-guided regularization technique. ``w/o Regularization'' indicates results without applying category-guided regularization technique.}
	\label{fig9}
\end{figure}

\textbf{Effectiveness of category-guided regularization technique.} Figure~\ref{fig9} presents the ablation results of employing the category-guided regularization technique. As shown in the figure, this technique effectively mitigates the Janus problem. For example, in the second column, the panda exhibits multiple faces, and the horse's ear position appears abnormal. By incorporating a multi-view consistent T2I model and category-guided prompts, this technique ensures spatial and semantic consistency across different viewpoints, leading to more coherent and realistic editing results.

\begin{figure}[h]
	\centering
	\includegraphics[width=0.9\columnwidth]{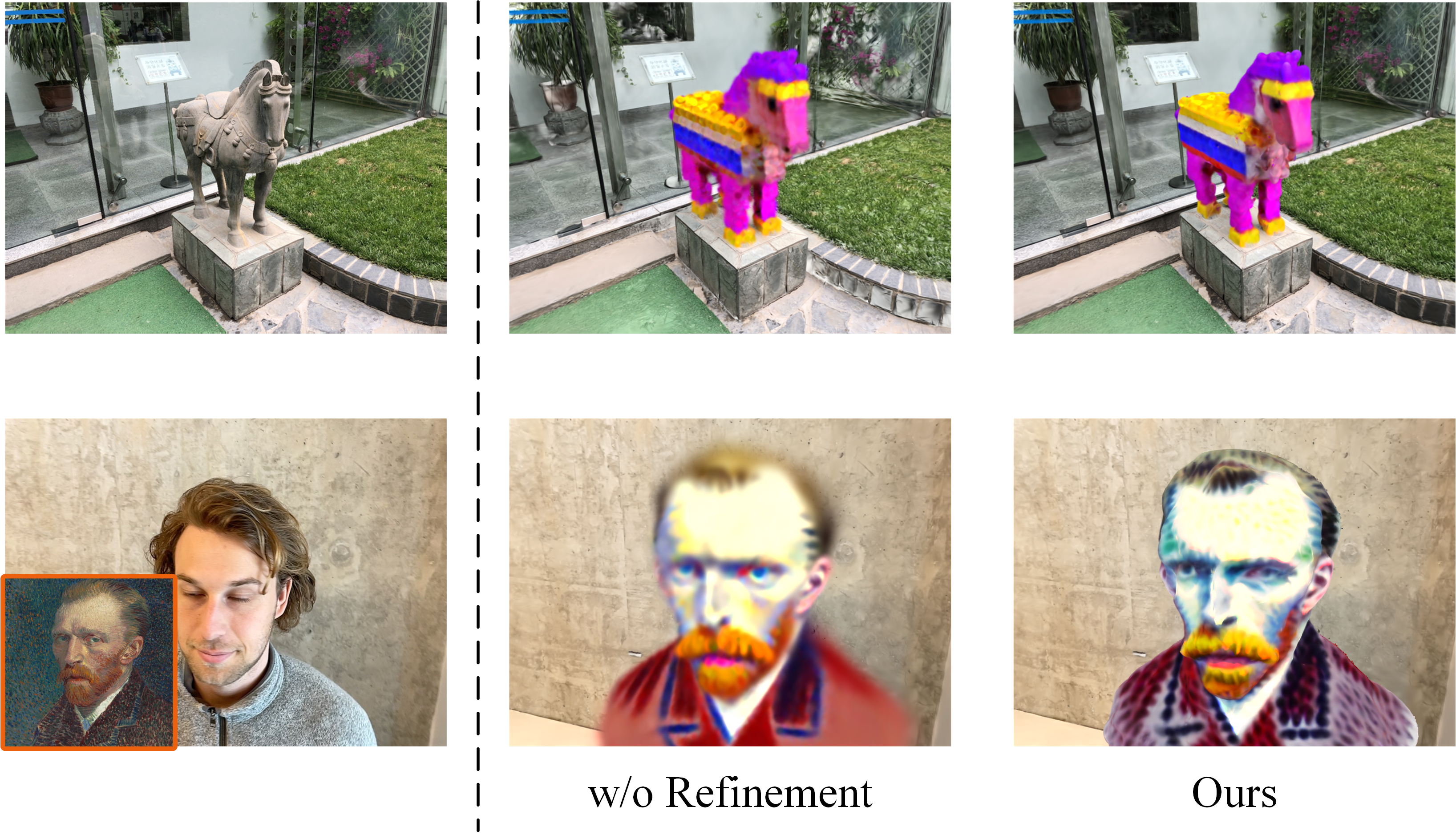}
	\caption{Ablation study of Texture Refinement. ``w/o Refinement'' indicates results without applying Texture Refinement.}
	\label{fig10}
\end{figure}

\textbf{Effectiveness of Texture Refinement.} Figure~\ref{fig10} and Table~\ref{tab3} present the ablation results of employing the Texture Refinement. When the Texture Refinement is applied, the texture and boundaries of the edited results become significantly clearer, and the edits using the reference image align more closely with the expected outcomes. For instance, in the first row, where the object is edited into a Lego horse, the results with Texture Refinement show noticeably sharper textures. Additionally, for tasks involving edits based on a reference image, the separation between the foreground and background is more distinct, and the results better resemble the reference image. The data in Table~\ref{tab3} strongly supports the qualitative analysis presented above. This improvement is attributed to the iterative refinement process, which enhances texture details and enforces closer adherence to the semantic and visual cues provided by the reference image.

\section{Conclusion and Limitations}\label{cl}
In this paper, we present GaussEdit, a novel approach for 3D scene editing driven by unified text and image prompts. GaussEdit offers significant improvements in both flexibility and precision, enabling intuitive and accurate scene modifications. Our contributions include a three-stage framework that utilizes 3D Gaussian Splatting for scene representation, an Adaptive Global-Local Optimization strategy to enhance editing consistency, a category-guided regularization technique to address the Janus Problem, and an advanced texture refinement method that ensures high-quality results.

Despite these strengths, GaussEdit has some limitations. One notable limitation is the reliance on coarse bounding box inputs to define the ROI. While convenient, this method struggles in complex scenes where bounding boxes may include unwanted elements. This issue becomes particularly evident with objects of irregular shapes or when objects are closely intertwined with their surroundings, potentially leading to inaccuracies at the boundaries of edited regions. Future research could explore more precise alternatives, such as employing segmentation methods based on point-cloud semantic segmentation, user-interactive selection via scribbles, or leveraging advanced neural methods to detect and isolate complex shapes more accurately. Such enhancements would likely result in more precise editing and improved user experience, especially for scenes with intricate details. Additionally, expanding GaussEdit to handle dynamic or deformable objects remains an interesting area for future work. Another issue is the sensitivity to hyperparameters, such as the CFG value and the learning rate of the Gaussian parameters. Currently, the quality of the editing results is affected by the choice of these hyperparameters. In future work, we plan to explore adaptive hyperparameter tuning strategies to mitigate this sensitivity and improve robustness.

\begin{table}[t]
	\centering
	\caption{Quantitative analysis for the Texture Refinement. ``w/o Refinement'' indicates results without applying Texture Refinement.}
	\label{tab3}
	\renewcommand\tabcolsep{4pt}
	\small
	\begin{tabular}{lcc}
		\toprule
		& CLIP$_{dir} \uparrow$   & DINO$_{sim} \uparrow$ \\
		\midrule
		w/o Refinement   & 26.53                   & 0.657                 \\
		Ours             & \textbf{28.11}          & \textbf{0.682}        \\
		\bottomrule
	\end{tabular}
\end{table}

\section*{Acknowledgments}\label{sec:acknowledgements}
This work is supported by the National Natural Science Foundation of China (62172356, 61872321), Zhejiang Provincial Natural Science Foundation of China (LZ25F020012),
the Ningbo Major Special Projects of the ``Science and Technology Innovation 2025'' (2020Z005, 2020Z007, 2021Z012, 2024Z122).

\ifCLASSOPTIONcaptionsoff
\newpage
\fi

\bibliographystyle{IEEEtran}
\bibliography{reference}



\begin{IEEEbiography}[{\includegraphics[width=1in,height=1.25in,clip,keepaspectratio]{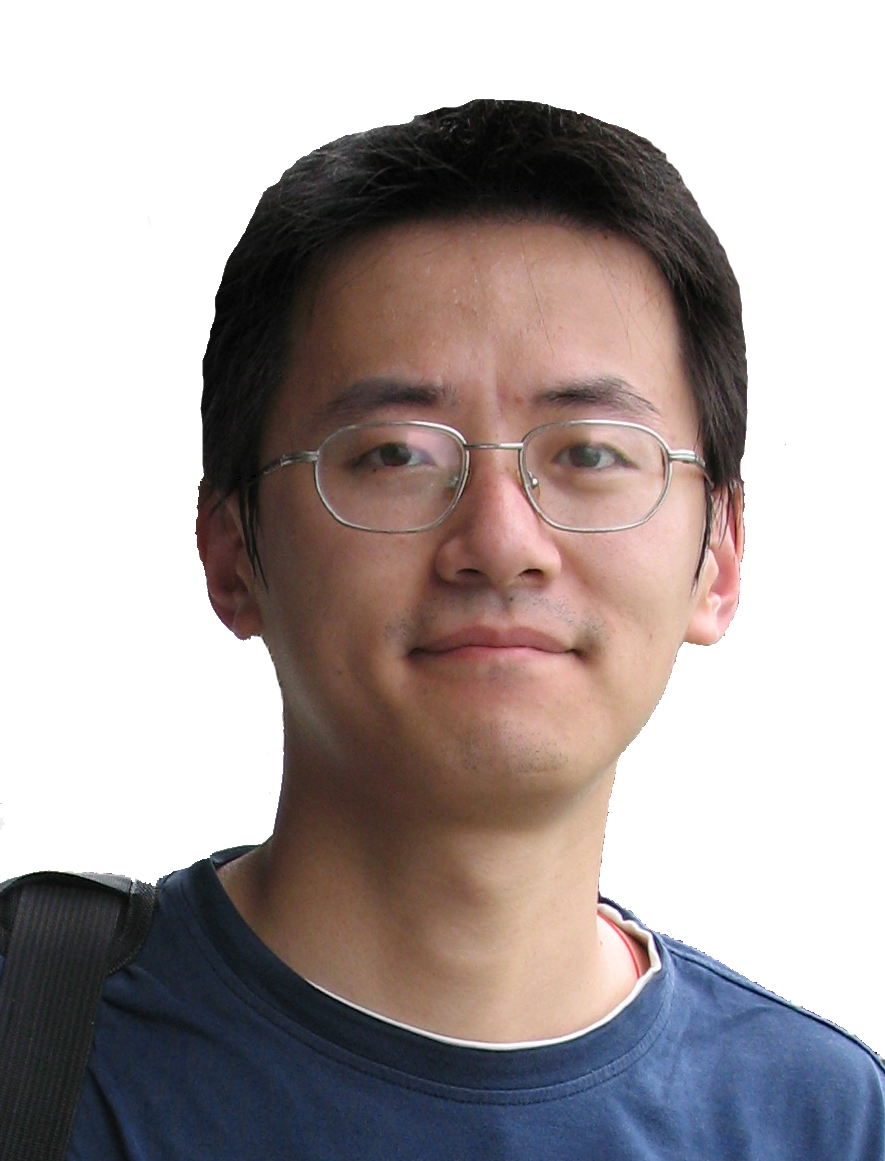}}]{Zhenyu Shu}
	got his Ph.D. degree in 2010 at Zhejiang University, China. He is now working as a full professor at NingboTech University. His research interests include computer graphics, digital geometry processing and machine learning. He has published over 30 papers in international conferences or journals.
\end{IEEEbiography}

\begin{IEEEbiography}[{\includegraphics[width=1in,height=1.25in,clip,keepaspectratio]{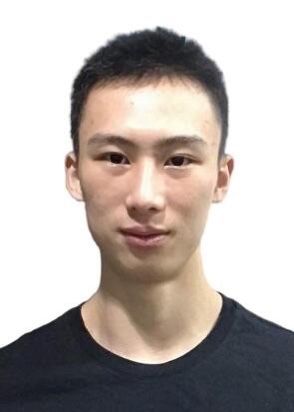}}]{Junlong Yu}
	is a graduate student of the School of Software Technology at Zhejiang University. His research interests include computer graphics and machine learning.
\end{IEEEbiography}

\begin{IEEEbiography}[{\includegraphics[width=1in,height=1.25in,clip,keepaspectratio]{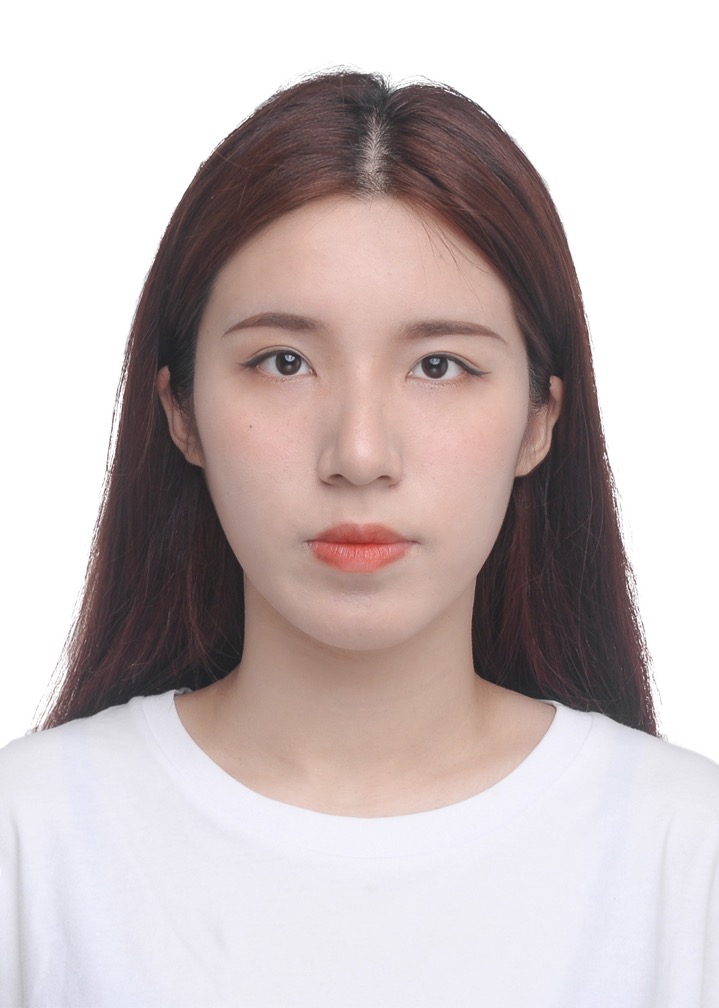}}]{Kai Chao}
	is a graduate student of the College of Management at Xi'an Jiaotong University. Her research interests include machine learning and AI management.
\end{IEEEbiography}

\begin{IEEEbiography}[{\includegraphics[width=1in,height=1.25in,clip,keepaspectratio]{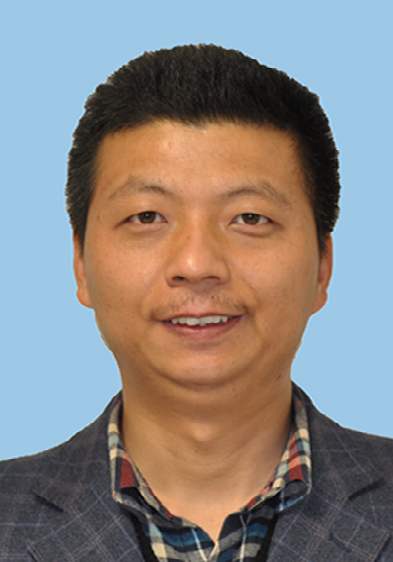}}]{Shiqing Xin}
	is an associate professor at the Faculty of School of Computer Science and Technology in Shandong University. He received his Ph.D. degree in applied mathematics at Zhejiang University in 2009. His research interests include computer graphics, computational geometry and 3D printing.
\end{IEEEbiography}

\begin{IEEEbiography}[{\includegraphics[width=1in,height=1.25in,clip,keepaspectratio]{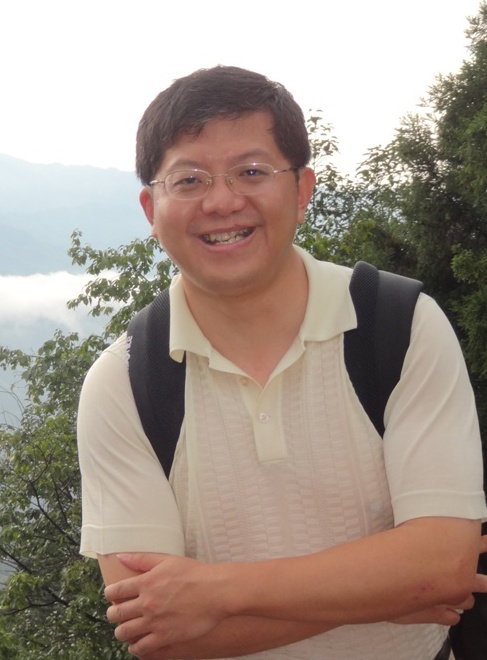}}]{Ligang Liu} received the BSc degree in 1996 and the Ph.D. degree in 2001 from Zhejiang University, China. He is a professor at the University of Science and Technology of China. Between 2001 and 2004, he was at Microsoft Research Asia. Then he was at Zhejiang University during 2004 and 2012. He paid an academic visit to Harvard University during 2009 and 2011. His research interests include geometric processing and image processing. He serves as the associated editors for journals of IEEE Transactions on Visualization and Computer Graphics, IEEE Computer Graphics and Applications, Computer Graphics Forum, Computer Aided Geometric Design, and The Visual Computer. His research works could be found at his research website: http://staff.ustc.edu.cn/lgliu
\end{IEEEbiography}

\vfill







\end{document}